\documentclass[preprint,overfull]{epl}

\title{Diblock copolymer ordering induced by patterned surfaces}
\shorttitle{Copolymer ordering induced by patterned surfaces}
\author{Yoav Tsori \and David Andelman}
\shortauthor{Y. Tsori \and D. Andelman}
\institute{
School of Physics and Astronomy\\
Raymond and Beverly Sackler Faculty of Exact Sciences\\
Tel Aviv University, 69978 Ramat Aviv, Israel
}
\date{12/12/2000}

\begin{document}

\maketitle

\begin{abstract}
 We use a Ginzburg-Landau free energy functional to
investigate diblock copolymer morphologies when the copolymer melt
interacts with one surface or is confined between two chemically
patterned surfaces. For temperatures above the order-disorder
transition a complete linear response description of the copolymer
melt is given, in terms of an arbitrary two-dimensional surface
pattern. The appearance of order in the direction parallel to the
surface is found as a result of the order in the perpendicular
direction. Below the order-disorder transition and in a thin-film
geometry, our procedure enables an analytic calculation of
distorted perpendicular and tilted lamellar phases in the presence
of uniform or striped surface fields.

\end{abstract}

Block copolymers (BCP) are macromolecules made up of two or more
chemically distinct
subunits, or blocks, covalently bonded together. The usual (macro-) phase
separation occurring for two immiscible species is not possible for
BCP because of
the covalent bond between the blocks.
The most studied BCP are the diblocks, for which the phase diagram
is quite well understood
\cite{B-F90,O-K86,Leibler80,B-F-SJPII97,F-H87,M-B96}, and
was found to
consist of disordered, lamellar, hexagonal and cubic micro-phases with
characteristic length scale usually in the range of dozens of nanometers.
The prevailing
morphologies depend on three parameters: the Flory parameter $\chi$,
characterizing the incompatibility between the two blocks, the
polymerization degree of the chain $N$, and
the fraction $f=N_A/N$
of the A block in a chain of $N=N_A+N_B$ monomers.
For high enough temperatures
the system is found in its disordered state. For symmetric melts
($f=1/2$) lowering the temperature (or equivalently, raising $\chi$) results in
a phase transition to the lamellar phase through the order-disorder
transition (ODT) point. For asymmetric melts ($f\neq 1/2$) the ordered phases
can have, in addition, hexagonal
(cylindrical) or cubic symmetries
\cite{B-F90,O-K86,Leibler80,B-F-SJPII97,F-H87,M-B96}.

The technological importance of copolymer thin-film is prominent in
diverse fields, such
as in fabrication of nanolithographic templates \cite{Chaikin97},
waveguides, anti-reflection coating for optical surfaces
\cite{M-SSCIENCE99} and dielectric mirrors \cite{fink98}.
Therefore, it is highly desired to acquire a better
understanding of the copolymer behavior in the presence of chemically
patterned surfaces.
Theoretical
\cite{Fredrickson87,T-F92,shull92,turnerPRL92,T-R-M94,matsenJCP97,M-MPRE96,P-BMM97,P-Muthu97-98,G-M-B00,P-WPRE99}
and experimental \cite{A-R-S-M-PRL89,M-R-A-S-M-PRL92,L-RMACRO94,L-RPRL96,M-RPRL97}
studies have been carried out in order to explore how the BCP film thickness,
the range and strength of the uniform surface
interactions, as well as the
bulk parameters $\chi$, $N$ and $f$, compete to produce a rich interfacial
behavior. Below the ODT,
these studies \cite{matsenJCP97,P-BMM97,G-M-B00} revealed
parallel, perpendicular  and mixed  lamellar phases, the apparent morphology
being the one that optimizes interaction with the surface and periodicity
frustration. Alignment of the lamellae can also be achieved through the
coupling of electric fields to local dielectric constant variations
\cite{M-RSCIENCE96}, or by shear \cite{P-BJPII97}.

Let us consider first a BCP melt in its disordered phase (above
the bulk ODT temperature) and confined by one or two flat,
chemically patterned surfaces. Although the  bulk BCP is disordered
above the ODT, there is an oscillatory decay of the A-B correlations
and surface induced ordering is quite complex. In the vicinity of
the ODT this ordering can become long range leading to a strong
effect. Defining the order parameter $\phi({\bm
r})\equiv\phi_A({\bm r})-f$ as the deviation of the A monomer
local concentration from its average, the free energy (in units of the
thermal energy $k_BT$) can be written as:

\begin{equation}\label{F}
F=\int\left\{\frac12\tau\phi^2+\frac12h\left[\left(\nabla^2+
q_0^2\right)\phi\right]^2 +\frac{u}{4!}\phi^4-\mu\phi\right\}{\rm
d}^3{\bm r}
\end{equation}
$d_0=2\pi/q_0$ is the fundamental periodicity in the system, and is
expressed by the polymer radius of gyration $R_g$, through
$q_0\simeq 1.95/R_g$. In addition,
$\tau=2\rho N\left(\chi_c-\chi\right)$,
$h=1.5\rho c^2R_g^2/q_0^2$ and $\mu$ is the chemical potential. The
second length scale in the system is determined by the ratio of
two parameters,
$\left(\tau/h\right)^{-1/4}\sim\left(N\chi_c-N\chi\right)^{-1/4}$,
and it characterizes the decay of surface induced modulations. The
Flory parameter $\chi$ measures the distance from the ODT point,
having the value $\chi_c\simeq 10.49/N$. Finally, $\rho=1/Na^3$ is
the chain density per unit volume, and $c$ and $u/\rho$ are
dimensionless constants of order unity \cite{B-F-SJPII97}.

This and similar types of free energy have been used successfully
to describe spatially modulated phases
\cite{Swift77,AndelmanSC95,mukamelPRE00}, with applications to
amphiphilic systems \cite{G-S90,G-ZPRA92}, diblock copolymers
\cite{Leibler80,B-F-SJPII97,F-H87,T-A-S00,N-A-SPRL97}, Langmuir
films \cite{A-B-J87} and magnetic (garnet) films \cite{G-D82}.
The free energy, Eq.~(\ref{F}), describes  a system in the
disordered phase having a uniform $\phi=0$ for $\chi<\chi_c$,
while  for $\chi>\chi_c$ (but in the vicinity of $\chi_c$), the
system is in the lamellar phase and is described  approximately by
a single $q$-mode $\phi=\phi_q\exp(i{\bm q_0}\cdot {\bm r})$.
The use of Eq.~(\ref{F}) limits us to a region of the
phase diagram close enough to the critical point where the expansion
in powers of $\phi$ and its derivatives is valid,
but not too close to it, because then critical fluctuations are
important \cite{B-F90}.

The presence of chemically heterogeneous surfaces is modeled by
adding a short-range surface interactions to the free energy,
\begin{equation}\label{Fs}
F_s=\int{\rm d^2}{\bm r_s}\left(\sigma({\bm r_s})\phi({\bm r_s}) +\tau_s\phi^2({\bm
r_s})\right)
\end{equation}
where the vector ${\bm r}={\bm r_s}$ defines the position of the
confining surfaces. The surface field $\sigma({\bm r_s})$ has an
arbitrary but fixed spatial variation and is coupled linearly to
the BCP surface concentration $\phi({\bm r_s})$. Preferential
adsorption of the A block is modeled by a constant $\sigma<0$
surface field, resulting in parallel-oriented layers (a
perpendicular orientation of the chains).  Control over the
magnitude of this surface field can be achieved by coating the
substrate with carefully prepared random copolymers
\cite{L-RPRL96,M-RPRL97}. If the pattern is spatially modulated,
$\sigma({\bm r_s})\neq 0$, then the A and B blocks are attracted
to different regions of the surface. The coefficient of the
$\phi^2$ term in Eq.~(\ref{Fs}) is taken to be a constant surface
correction to the Flory parameter $\chi$
\cite{Fredrickson87,T-F92}. A positive $\tau_s$ coefficient
corresponds to a suppression of surface segregation of the A and
B monomers.

We consider first the semi-infinite system of polymer melt in
contact with a single flat surface given by ${\bm r_s}=(x,y=0,z)$.
The ordering effect is expressed in terms of the correction to the
order parameter $\delta\phi({\bm r})=\phi-\phi_b$ about its bulk
value $\phi_b$. This correction vanishes in the bulk,
$\lim_{y\rightarrow \infty}\delta\phi=0$, because of the finite
range of surface interactions. Moreover, the correction should
preserve the condition of fixed A/B ratio, namely $\int
\delta\phi({\bf r}){\rm d}^3r=0$. The free energy can be expanded
about the bulk value $F[\phi_b]$ to second order in $\delta
\phi$: $F=F[\phi_b]+\Delta F$, with

\begin{eqnarray}\label{DeltaF}
\Delta F &=&\int \left\{\left[(\tau+hq_0^4)\phi_b+
hq_0^2\nabla^2\phi_b+
\frac{1}{6}u\phi_b^3-\mu\right]\delta\phi \right. \nonumber\\
&+&\left. \frac12(\tau+\frac12 u\phi_b^2)\delta\phi^2+
\frac12h\left[\left(q_0^2+\nabla^2\right)\delta\phi\right]^2\right\}
{\rm d^3}{\bm r}\nonumber\\
&+&\int \{\sigma(x,z)\delta\phi_s+
\tau_s\left(2\phi_b\delta\phi_s+\delta\phi_s^2\right)\}{\rm d}x{\rm d}z
\end{eqnarray}
$\delta\phi_s=\delta\phi(x,0,z)$ is the surface value of
$\delta\phi$. The surface chemical pattern $\sigma({\bm
r_s})=\sigma(x,z)$ can be decomposed in terms of its ${q}$-modes
$\sigma(x,z)=\sum_{\bm q}\sigma_{\bm
q}\exp[i\left(q_xx+q_zz\right)]$, where ${\bm q}\equiv(q_x,q_z)$,
and $\sigma_{\bm q}$ is the mode amplitude.
Similarly, the correction field is $\delta\phi(x,y,z)= \\
\sum_{\bm q}\delta\phi_{\bm q}(y)\exp[i\left(q_xx+q_zz\right)]$,
and is substituted in Eq.~(\ref{DeltaF}). We will first consider
a system found above the ODT temperature, in the disordered
phase. In this case $\phi_b=0$, and the $x$ and $z$ integration
of the free energy, Eq.~(\ref{DeltaF}), is carried out
explicitly. Then, applying a variational principle with respect
to $\delta\phi_{\bm q}$ results in the following differential
equation:
\begin{eqnarray}\label{EL-fq}
\left(\tau/h+\left(q^2-q_0^2\right)^2\right)\delta\phi_{\bm q}
+2(q_0^2-q^2)\delta\phi_{\bm q}^{\prime\prime}+\delta\phi_{\bm
q}^{\prime\prime\prime\prime}=0
\end{eqnarray}
This ordinary differential equation is linear and of fourth order.
In the semi-infinite geometry, $y>0$, the solution to
Eq.~(\ref{EL-fq}) has an exponential form $\delta\phi_{\bm q}(y)=A_{\bm q}\exp(-k_{\bm
q}y)+B_{\bm q}\exp(-k^*_{\bm q}y)\label{fq}$,
where $k_{\bm q}$ is given by
\begin{eqnarray}
k_{\bm q}^2 &=&q^2-q_0^2+i\sqrt{\tau/h}\nonumber\\
&=&q^2-q_0^2+i\alpha\left(N\chi_c-N\chi\right)^{1/2}\label{kq}
\end{eqnarray}
Out of the four complex roots for $k_{\bm q}$, the two with ${\rm
Re}(k_q)<0$ diverge at $y\rightarrow\infty$ and are discarded.
From the requirement of fixed A/B ratio it follows that the
chemical potential is $\mu=0$. For numerical purposes and in all
plots we set the fundamental monomer length as $a=1$, and choose
in Eq. (\ref{F}) $c=u/\rho=1$  to give $\alpha\simeq0.59q_0^2$.
The values of $\xi_q=1/{\rm Re}(k_{\bm q})$ and $\lambda_q=1/{\rm
Im}(k_{\bm q})$ correspond to the exponential decay and
oscillation lengths of the ${q}$-modes, respectively. For fixed
$\chi$, $\xi_q$ decreases and $\lambda_q$ increases with
increasing $q$. Similar behavior was found by Petera and
Muthukumar \cite{P-Muthu97-98}, using a different free energy
functional \cite{O-K86}. Close to the ODT (but within the range
of validity of the model), and for ${q}$-modes such that $q>q_0$
we find finite $\xi_q$ and $\lambda_q\sim(\chi_c-\chi)^{-1/2}$.
However, all $q$-modes in the band $0<q<q_0$ are equally
``active'', i.e., these modes decay to zero very slowly in the
vicinity of the ODT as $y\rightarrow\infty$:
$\xi_q\sim(\chi_c-\chi)^{-1/2}$ and $\lambda_q$ is finite.
Therefore, the propagation of the surface imprint (pattern) of
$q$-modes with $q<q_0$ into the bulk can persist to long
distances, in contrast to surface patterns with $q>q_0$ which
persist only close to the surface. The $q=q_0$ mode has both
lengths $\xi_q, \lambda_q$ diverging as $(\chi_c-\chi)^{-1/4}$
for $\chi\rightarrow\chi_c$.

The boundary conditions at $y=0$
for $\delta\phi_{\bm q}$ are
\begin{eqnarray}\label{bc1}
\delta\phi_{\bm q}^{\prime\prime}(0)+\left(q_0^2-q^2\right)
\delta\phi_{\bm q}(0)&=&0\\ \label{bc2}
\sigma_{\bm q}/h+2\tau_s\delta\phi_{\bm
q}(0)/h+\left(q_0^2-q^2\right)\delta\phi_{\bm
q}^{\prime}(0)+\delta\phi_{\bm q}^{\prime\prime\prime}(0)
&=&0
\end{eqnarray}
The amplitude $A_q$ is found to be
$A_q=-\sigma_q\left(4\tau_s+2{\rm Im}(k_q)\sqrt{\tau
h}\right)^{-1}$. Thus, in cases
where the surface orders at the same temperature
as the bulk, $\tau_s=0$, the copolymer response
diverges upon approaching the critical point as
$\left(N\chi_c-N\chi\right)^{-1/2}$.

With our method, any two-dimensional chemical pattern $\sigma(x,z)$
can be
modeled.
For surface feature size larger than $d_0$, the
characteristic copolymer length, the melt can propagate the chemical
surface
pattern into the bulk.  This is
clearly seen in Fig.~1,
where the surface pattern separating A regions and B regions is chosen
arbitrarily to have the shape of
the letter `A' of size $15d_0 \times 15d_0$.
The order propagating perpendicular to the surface, Fig.~1, also induces
order in a parallel direction, where lamellae appear oriented along the long edges
of the letter. This ordering decays exponentially
in the lateral direction, with decay length determined by the surface $q$-modes.
The shading of the contours in (b) is approximately
opposite to that of (a) and (c) indicating a reversal of the
pattern; this is expected for two planes
separated by a half-integer number of $d_0$, a distance for which an
A$\leftrightarrow$B interchange of monomers occurs. The distance from
the
surface at which the pattern completely fades out depends on the surface
Fourier components: each $q$-mode decays as dictated by
Eq.~(\ref{kq}), and the distant image is a superposition of all modes. In (c)
the pattern is almost washed out
at a distance $12 d_0$ from the surface.
Figure~2 (a) is similar to Fig.~1 (b) (with $y=0.5d_0$), but we scaled
down the `A' pattern so that its size is $4d_0 \times 4d_0$. For
surface feature size smaller than $d_0$, the melt cannot follow the
surface pattern, and the morphology is blurred even very close to
the surface. As one goes deeper into the disordered phase (lowering
$\chi$), the lamellar features are less apparent, as is seen in
Fig.~2 (b) for which $\chi N=9.5$ and all other parameter are taken as
in Fig.~1 (b).

Our treatment can be generalized to handle a thin BCP film of
thickness $2L$ confined between two chemically patterned
surfaces. The two corresponding surface  patterns (located at
$y=\pm L$) are defined by two surface fields
$\sigma^{\pm}(x,z)=\sum_{\bm q}\sigma^{\pm}_{\bm q}
\exp{[i\left(q_xx+q_zz\right)]}$. The form of the response
function $\delta\phi$ is required  now to satisfy four boundary
conditions (similar to  Eqs.~(\ref{bc1}) and (\ref{bc2})), two on
each surface.
For very large separation $2L$, the melt orders close to the
$y=\pm L$ surfaces, while the middle of the film, $y\approx 0$,
is in its disordered
state. As
the inter-surface separation decreases, the ordered layers close to the two
surfaces start to overlap.

We briefly mention results for a thin-film BCP below its bulk ODT
temperature \cite{tobepub}, where the two surface fields
$\sigma^{\pm}$ at $y=\pm L$ are taken to be uniform. Their values
can be positive or negative depending on their preference to the A
and B blocks. The free energy Eq.~(\ref{DeltaF}) expresses a
correction to a lamellar phase perpendicular to the surface. The
correction field $\delta\phi(y)$ has the form ${\rm Re}[
A_0\exp(-k_0y)+B_0\exp(-k_0^*y)+const.]$. A similar perturbed
lamellar phase was found in the strong or intermediate segregation
regimes \cite{matsenJCP97,G-M-B00}. However, close to the ODT the
lamellar deformation  near the surface changes the free energy
considerably. This is seen in Fig.~3 (a) for the symmetric case,
$\sigma^+=\sigma^-$. The B-monomers (in black) are attracted to both
surfaces. The oscillatory excess free energy $\Delta F<0$ is shown
in (b) as a function of the separation $2L$ between the surfaces,
indicating that the deformation $\delta\phi$ indeed lowers the
total free energy. In (c) a semi-infinite system is shown, with one
sinusoidally patterned surface of period $d$ at $y=0$. The lamellae
appear tilted with an angle $\theta=\arcsin(d_0/d)$ with respect to
the surface, in agreement with Ref.~\cite{P-Muthu97-98}. Close to the surface
the deformation may be very large.

In summary, we show a simple method to find the surface effects
in confined diblock copolymers. The free energy is expanded
to second order around a bulk (ordered or disordered) phase.
Above the ODT we obtain
a master equation for the response of copolymer
melt to an arbitrary two-dimensional surface pattern.
These two-dimensional patterns have not been considered previously
\cite{P-Muthu97-98} and show a rich behavior as
the BCP order parameter replicates (to some extent) the surface
pattern.
In order to transfer a pattern from the surface to the bulk, or
from one surface to another via a BCP film, it is important to
control the thermodynamic conditions, and in particular, how
close we are to the ODT. Two opposite trends should be taken into
account. On one hand, far from the ODT, the required surface
fields may be large. On the other hand, close to the ODT, high
$q$-modes decay much faster than low $q$-modes, resulting in a
distorted pattern. Below the ODT we find the profile and energy
of the distorted  perpendicular lamellar phase as function of the
surface fields and film thickness. When the surfaces are
chemically modulated tilted lamellar phases are stabilized
\cite{tobepub}.

Surface behavior of BCP or thin films of BCP confined between two
surfaces may be used in many applications. For example, surface
patterns can be used to generate a desired morphology inside a
BCP film. Then, one of the two components is removed leaving
behind a pre-designed porous material with controlled morphology.

\acknowledgments We would like to thank S. Herminghaus,
G. Krausch, M. Muthukumar, R. Netz, G. Reiter, T. Russell, M.
Schick, M. Schwartz and U. Steiner for useful comments and
discussions. Partial support from the U.S.-Israel Binational
Foundation (B.S.F.) under grant No. 98-00429 and the Israel Science
Foundation funded by the Israel Academy of Sciences and Humanities
--- Centers of Excellence Program is gratefully acknowledged.

\newpage

\begin{figure}
\onefigure[scale=0.7]{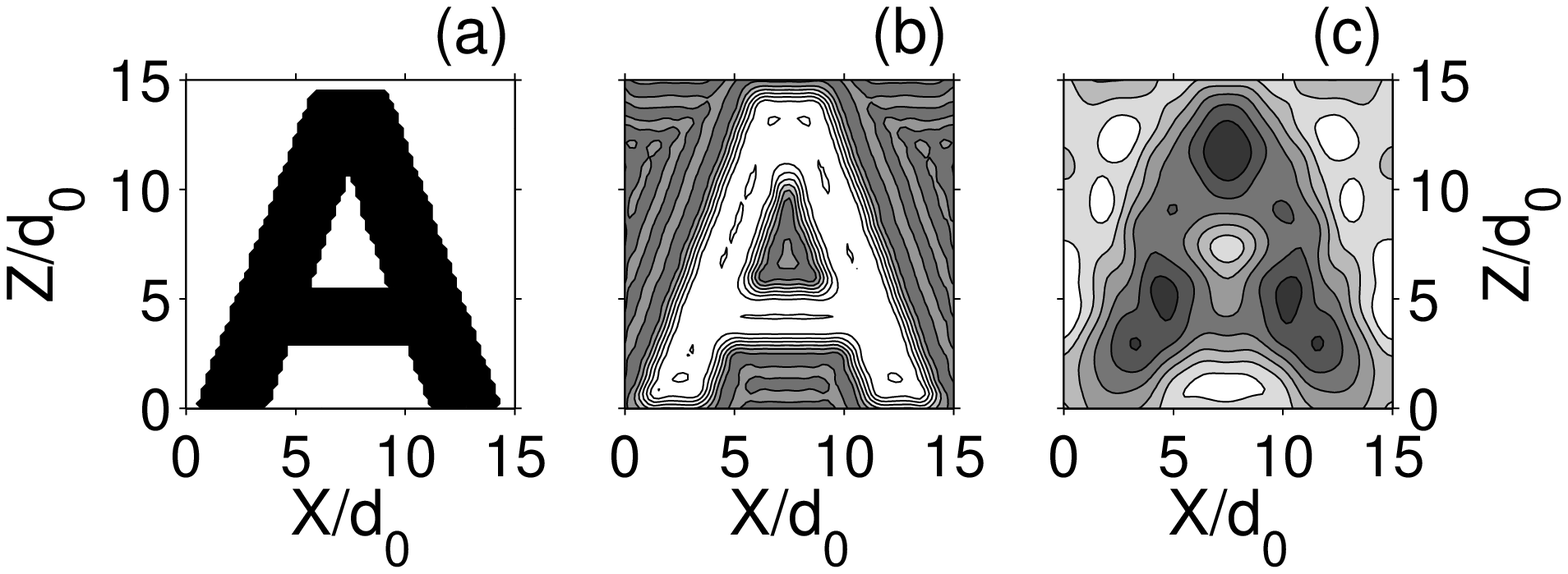}
\caption{Contour plots of the diblock copolymer order parameter
  depicting
slices parallel to the $y=0$ surface. (a) The surface pattern at
$y=0$, chosen in the shape of
the `A' letter  of size $15d_0 \times 15d_0$. Inside the letter
`A' $\sigma=1$, while outside $\sigma=0$. In (b) the BCP morphology is
shown for $y=0.5d_0$, and in (c) for $y=12d_0$. For planes separated by
a half integer number of
$d_0$, an A$\leftrightarrow$B interchange of monomers occurs
(compare (b) to (a) and (c)). The order propagating in the
perpendicular direction induces order in a direction parallel to
the surface. This is clearly seen in (b) where lamellae form
parallel to the edges of the `A' letter.
 The Flory parameter is $\chi
N=10.3$ and seven gray scales
are used to maximize the image contrast.
All lengths are scaled by the lamellar period $d_0$,
and the parameters chosen in Eq. (\ref{F}) are: $N=1000$, $R_g^2=\frac16
Na^2$, $a=c=u/\rho=1$ and
$\tau_s=0.03$. These values are used in Figs. 2-3 as well.
}
\label{fig1}
\end{figure}

\begin{figure}
\onefigure[scale=0.7]{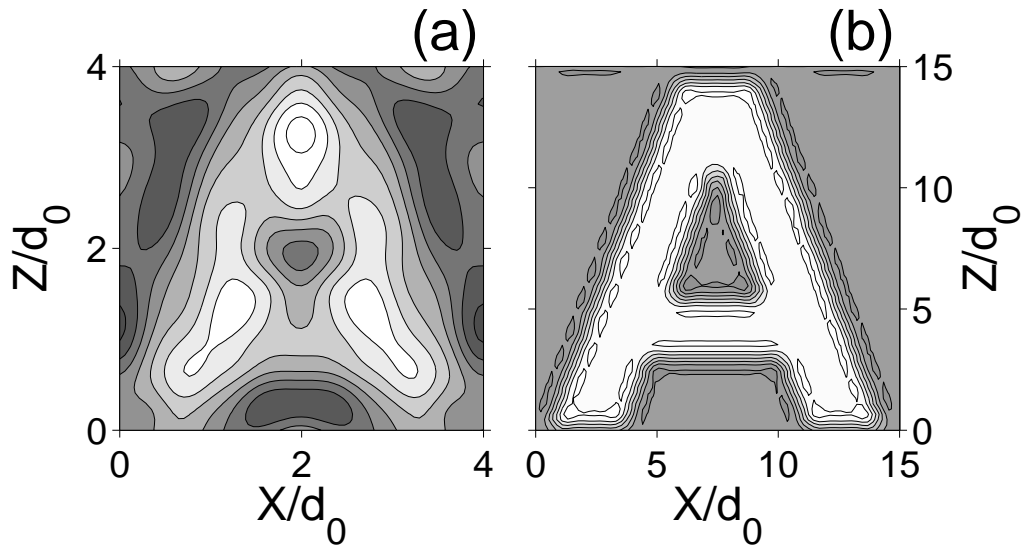}
\caption{Contour plots as in Fig.~1 (b) ($y=0.5d_0$),
but in (a) the surface pattern is
reduced to a size of
$4d_0\times 4d_0$.
Note that the surface cannot induce a bulk ordering when its
pattern size is close to $d_0$ or smaller. In (b)
the temperature is higher,
$\chi N=9.5$, and the lamellar features along the
letter are less prominent than in Fig.~1 (b).
}
\label{fig2}
\end{figure}

\begin{figure}
\onefigure[scale=0.8]{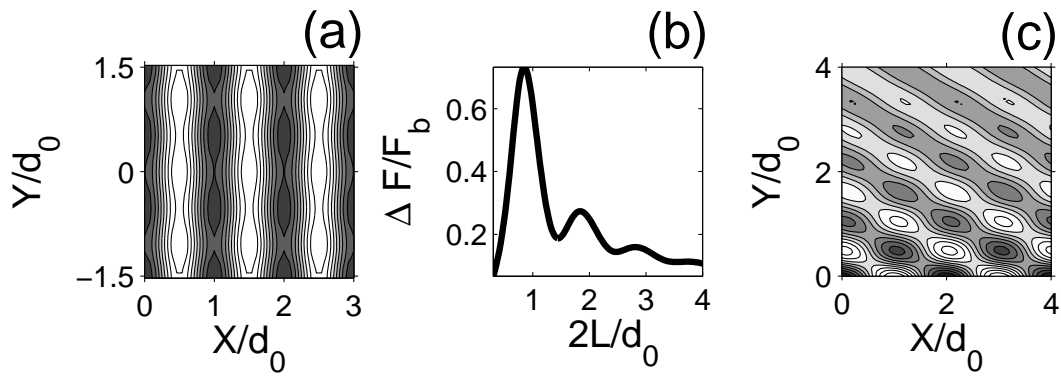} \caption{(a) Perpendicular
lamellar phase between two (top and bottom) surfaces below the
ODT ($\chi N=10.7$).
 The two surfaces at $y=\pm L$ have
equal surface fields, $\sigma^{+}=\sigma^{-}=0.07$, preferring the
B-block (in black). (b) The oscillatory character of the
correction free energy $\Delta F<0$ (Eq. \ref{DeltaF}) of the
symmetric system in (a), as a function of surface separation $2L$.
$\Delta F$ is divided by the bulk lamellar free energy $F_b<0$,
and it decays to zero for $L\to \infty$. (c) Tilted lamellar
phase occurring for a single sinusoidally patterned surface of
period $d>d_0$ at $y=0$. The lamellae gain interfacial energy by
overlapping with the surface pattern. The Flory parameter is
$\chi N=11$ and the surface periodicity is chosen as $d/d_0=2$
giving a tilt angle of $\theta=30^\circ$.} \label{fig3}
\end{figure}

\end{document}